\documentclass[11pt]{article}

\usepackage{latexsym,amsmath,amssymb}

\numberwithin{equation}{section}

\newcommand{\be}{\begin{equation}}
\newcommand{\ee}{\end{equation}}

\newcommand{\bea}{\begin{eqnarray}}
\newcommand{\eea}{\end{eqnarray}}

\newcommand{\ns}{\normalsize}
\newcommand{\pt}{\partial}

\def\a{\alpha}
\def\b{\beta}
\def\g{\gamma}

\def\d{\delta}
\def\e{\epsilon}

\def\f{\phi}

\def\z{\psi}

\def\m{\mu}
\def\n{\nu}

\def\p{\pi}

\def\r{\rho}
\def\s{\sigma}
\def\t{\tau}

\def\z{\zeta}

\def\L{\Lambda}

\def\cH{{\mathcal H}}

\def\cL{{\mathcal L}}
\def\cV{{\mathcal V}}

\def\ZM{{\mathbb Z}}

\def\bbe{\boldsymbol{\beta}}
\def\bph{\boldsymbol{\phi}}
\def\bla{\boldsymbol{\lambda}}
\def\btau{\boldsymbol{\tau}}

\begin{document}

%%%%%%%%%%%%%%%%%%%%%%%%%%%%%%%%%%%%%%%%%%%%%%%%%%%%%%%%%%%%%%%%%%%%%

\begin{titlepage}

\vspace{-3cm}

\title{
   \hfill{\ns SUSX-TH/03-010\\}
   \hfill{\ns hep-th/0308195\\}
   \vskip 1cm
   {\Large Rolling $G_2$ Moduli}\\}
   \setcounter{footnote}{0}
\author{
{\ns\large Andr\'e Lukas\footnote{email: a.lukas@sussex.ac.uk}
  \setcounter{footnote}{3}
  and Stephen Morris\footnote{email: s.morris@sussex.ac.uk}} \\[0.8em]
   {\it\ns Department of Physics and Astronomy, University of Sussex}
   \\[-0.2em]
   {\ns Falmer, Brighton BN1 9QH, UK} \\[0.2em] }
\date{}

\maketitle

\begin{abstract}\noindent
We study the time evolution of freely rolling moduli in the context
of M-theory on a $G_2$ manifold. This free evolution approximates
the correct dynamics of the system at sufficiently large values of
the moduli when effects from non-perturbative potentials and flux
are negligible. Moduli fall into two classes, namely bulk moduli
and blow-up moduli. We obtain a number of non-trivial solutions
for the time-evolution of these moduli. As a generic feature, we
find the blow-up moduli always expand asymptotically at early
and late time.
\end{abstract}

\thispagestyle{empty}

\end{titlepage}

%%%%%%%%%%%%%%%%%%%%%%%%%%%%%%%%%%%%%%%%%%%%%%%%%%%%%%%%%%%%%%%%%%%%%%%

\section{Introduction}

Seven-dimensional manifolds with holonomy $G_2$ provide the
general framework for relating M-theory to four-dimensional models
with $N=1$ supersymmetry. Moreover, it has been shown that for
certain singular limits of such manifolds four-dimensional models
with phenomenologically interesting properties can be
obtained~\cite{Papadopoulos:1995da}--\cite{Acharya:2002kv}.

Given these promising features it seems worthwhile to have a closer
look at the cosmology of such $G_2$ compactifications of M-theory.
More specifically, in this paper, we will be interested in
the time-evolution of $G_2$ moduli. This problem has not previously
been analysed since the four-dimensional effective theories,
specifically the kinetic terms, for these fields were unknown.
However, in a recent paper~\cite{Lukas:2003dn}, the four-dimensional
moduli K\"ahler potential for M-theory on a $G_2$ manifold
has been computed explicitly and our analysis will be based on
these results.

\vspace{0.4cm}

The $G_2$ manifold considered in Ref.~\cite{Lukas:2003dn} is based on the
compact examples due to Joyce~\cite{joyce1}--\cite{joyceb},
constructed by blowing-up
a seven-dimensional $G_2$ orbifold. The moduli for such a $G_2$
manifold split into two groups, namely the bulk moduli (related
to the radii of the underlying seven-torus) and the blow-up moduli
(related to the volume of the blow-ups). The K\"ahler potential
of Ref.~\cite{Lukas:2003dn} was obtained in the usual large-radius supergravity
approximation and for blow-up volumes small compared
to the total volume of the space. In the limit of a vanishing
blow-up modulus the space develops a singularity (which
corresponds to a singularity of the underlying orbifold)
leading to new fields in the low-energy theory. It is singular
limits such as this which are essential for a phenomenologically
interesting low-energy theory.

\vspace{0.4cm}

In this paper, we will focus on the simple case of free moduli
evolution based on the K\"ahler potential of
Ref.~\cite{Lukas:2003dn}. That is we will neglect the effect of
possible non-perturbative potentials and flux. Hence our results
constitute a good approximation at large moduli values where
effects of a non-trivial moduli (super)-potential can be
neglected. Particular consideration is given to the evolution of
the blow-up moduli; particularly as to whether they may contract,
thereby approaching one of the phenomenologically interesting
singular limits.

\vspace{0.4cm}

Let us summarise our main results. Starting from the given
K\"ahler potential and consistently truncating off the axions in
the chiral moduli multiplets, we show that, after a suitable
reparametrisation of fields, the sigma-model metric becomes
independent of the blow-up moduli. As a result, these fields can
be integrated out and the problem reduces to one of studying the
evolution of the bulk moduli in an effective potential. Based on
this approach, we show that a number of explicit analytic
solutions can be obtained using methods from Toda theory. For all
these solutions, we find the relative volume of the blow-up moduli
increases asymptotically at early and late times. As a
consequence, the small blow-up approximation underlying the
K\"ahler potential breaks down after a finite proper time. This
feature can also be intuitively understood from the properties of
the effective bulk-moduli potential and can, therefore, be viewed
as generic. Hence, once moduli are large it appears to be
difficult to evolve towards a singular state of the internal
space. This suggests that a successful cosmological evolution
leading to phenomenologically interesting singular $G_2$ spaces
should stabilise moduli at small values from the outset.

%%%%%%%%%%%%%%%%%%%%%%%%%%%%%%%%%%%%%%%%%%%%%%%%%%%%%%%%%%%%%%%%%%%%%%%%%%%%%

\section{K\"ahler potential for $G_2$ moduli}

In Ref.~\cite{Lukas:2003dn} the moduli K\"ahler potential has been
calculated for a specific manifold with $G_2$ holonomy constructed
from a $G_2$ orbifold by blowing up the orbifold singularities.
The chiral moduli multiplets for this manifold split into two
categories, namely the bulk moduli $T^A$, where $A,B,C,\dots =
1,\dots ,7$, which encode the radii of the seven-torus $T^7$
underlying the orbifold, and the blow-up moduli $U^i$, which
measure the volume of the blow-ups. The general structure of this
K\"ahler potential is
\begin{equation}
 K = -\sum_{A=1}^7\ln\left(T^A+\bar{T}^A\right) + \sum_iF_i(T^A,\bar{T}^A)
     \left( U^i+\bar{U}^i\right)^2+c\; , \label{K}
\end{equation}
where the functions $F_i$ are given by
\begin{equation}
 F_i = \frac{8}{(T^{A_i}+\bar{T}^{A_i})(T^{B_i}+\bar{T}^{B_i})}
\end{equation}
and $c$ is a constant.
Here, $A_i$ and $B_i$ specify the two particular bulk moduli by which
the blow-up modulus $i$ is divided in the above K\"ahler potential.
The values for these indices reflect the underlying structure of the
orbifold and they can be conveniently encoded in constant seven-dimensional
vectors ${\rm p}_i$ by writing
\begin{equation}
 F_i = 2e^{-{\bf p}\cdot\bbe}
\end{equation}
where the new fields $\bbe = (\b_A)$ are defined by~\footnote{Here and in
the following bold-face symbols denote seven-dimensional vectors.}
\begin{equation}
 {\rm Re}(T^A)=e^{\b_A}\; . \label{beta}
\end{equation}
We expect this form of the K\"ahler potential to apply to a wide
class of $G_2$ manifolds constructed by blowing up orbifolds, with
the specifics of each example encoded in the vectors ${\bf p}_i$.

In the following we will, for concreteness, focus on the
particular example of Ref.~\cite{Lukas:2003dn}, which is based on
the orbifold $T^7/\ZM_2^3$. The blow-ups for this particular
manifold can be labeled by a triple $(i)=(\t ,n,a)$ of indices
where $\t =\a ,\b ,\g$ indicates the type, that is, under which of
the three $\ZM_2$ orbifolding symmetries the associated fixed
point remains invariant, $n=1,2,3,4$ labels the fixed points of
equal type and $a=1,2,3$ is an index associated to each fixed
point that describes the orientation of the blow-up relative to
the bulk. Hence, there are $12$ fixed points, labeled by $(\t
,n)$, each with three associated blow-up moduli and, consequently,
$36$ blow-up moduli in total. For simplicity of notation, we will
use the single index $i$ to label the blow-ups whenever possible
and only split into the triple $(\t ,n,a)$ when required. It turns
out that the vectors ${\bf p}_i$ for this case only depend on the
type $\t$ and the orientation $a$. The resulting nine vectors are
given in Table~1.
\begin{table}
 \begin{center}
 \begin{tabular}{|l|l|l|l|}
  \hline
  ${\bf p}_{(\t ,a)}$&$a=1$&$a=2$&$a=3$\\\hline
  $\t = \a$&$(1,0,0,0,0,1,0)$&$(0,1,0,0,1,0,0)$&$(0,0,1,1,0,0,0)$\\\hline
  $\t = \b$&$(1,0,0,0,0,0,1)$&$(0,0,1,0,1,0,0)$&$(0,1,0,1,0,0,0)$\\\hline
  $\t = \g$&$(1,0,0,1,0,0,0)$&$(0,0,1,0,0,1,0)$&$(0,1,0,0,0,0,1)$\\\hline
\end{tabular}
 \caption{Values of vectors ${\bf p}_{(\t ,a)}$ which define the
          moduli K\"ahler potential for the $G_2$ manifold based on
          $T^7/\ZM_2^3$.}
 \end{center}
\end{table}

\vspace{0.4cm}

It will be useful for the subsequent discussion to have an
interpretation for the moduli and the various parts of the K\"ahler
potential~\eqref{K} in terms of the geometry of the underlying $G_2$
manifold. To this end, one notes that the moduli K\"ahler potential
for M-theory on a $G_2$ manifold is generally related to the volume
$\cV$ of this manifold by~\cite{Beasley:2002db}
\begin{equation}
 K = -3\ln\left(\frac{\cV}{2\p^2}\right)\; . \label{KV}
\end{equation}
The first, $T^A$--dependent, part of the K\"ahler
potential~\eqref{K} then corresponds to the volume $\cV_0$ of the
orbifold, while the second, $U^i$--dependent, part measures the
reduction of this volume due to the blow-ups. For the orbifold
volume $\cV_0$ one easily finds from Eqs.~\eqref{KV} and
\eqref{beta} that
\begin{equation}
 \cV_0 = \frac{1}{8}\exp\left(\frac{1}{3}\sum_A\beta_A\right)\; ,
 \label{V0}
\end{equation}
where we have used the value $c=6\ln (8\p )+\ln (2)$ found in Ref.~\cite{Lukas:2003dn}.
From a similar calculation one finds that the fraction $\e_{(\t ,n)}$ by
which the blow-up $(\t ,n)$ reduces the orbifold volume $V_0$ is given by
\begin{equation}
 \e_{(\t ,n)} = \frac{8}{3}\sum_{a=1}^3e^{-{\bf p}_{(\t ,a)}\cdot\bbe}\,
                u_{(\t ,n,a)}^2\; ,
\end{equation}
where we have introduced the real parts
\begin{equation}
{\rm Re}(U^{i})=u_i
\end{equation}
of the blow-up moduli.

\vspace{0.4cm}

The K\"ahler potential~\eqref{K} relies on two approximations so
we should discuss its range of validity in moduli space. First, we
require all moduli to be larger than one, that is
\begin{equation}
 T^A\gg 1\; ,\qquad  U^i\gg 1\; ,\label{app1}
\end{equation}
for the supergravity approximation underlying the calculation of $K$ to be valid.
Second, the K\"ahler potential has been calculated to leading (quadratic)
order in $U^i/T^A$ and terms of order four in these ratios have been
neglected. Consequently, application of the K\"ahler potential~\eqref{K}
should be confined to the region of moduli space where the ratios
$U^i/T^A$ are smaller than one, or more precisely, where
\begin{equation}
 \e_{(\t ,n)} \ll 1\; . \label{app2}
\end{equation}
From the above interpretation of $\e_{(\t ,n)}$ this means that
the volume taken away by the blow-ups should be small compared to
the volume of the orbifold. In subsequent calculations, we will
consistently apply this approximation in that we neglect higher
order terms in $\e_{(\t ,n)}$.

%%%%%%%%%%%%%%%%%%%%%%%%%%%%%%%%%%%%%%%%%%%%%%%%%%%%%%%%%%%%%%%%%%%%%%%%%%%%%%%

\section{A simple form of the Lagrangian}

The K\"ahler potential described in the previous section depends on $43$
chiral superfields. We, therefore, have $43$ real moduli fields from the
real parts of theses superfields, which are associated to
the geometry of the $G_2$ manifold, plus $43$ axions from the imaginary
parts. It is clear that the axions can be consistently set to constants
and for simplicity this is what we will do in the following.
We are, hence, left with the real scalar fields $\b_A$ and $u_i$.

However, their kinetic terms, as computed from Eq.~\eqref{K}, are
still fairly complicated. A substantial simplification can be
achieved by introducing a new set of fields $\f_A$ and $z_i$
defined by the Ansatz
\begin{eqnarray}
 \b_A &=& \f_A+\sum_i c_{iA}e^{{\bf q}_i\cdot\bph}z_i^2 \\
 u_i &=& e^{{\bf s}_i\cdot\bph}\; .
\end{eqnarray}
Here, ${\bf c}_i=(c_{iA})$, ${\bf q}_i=(q_{Ai})$ and ${\bf
s}_i=(s_{Ai})$ are constants to be determined shortly. The kinetic
terms for $\b_A$ and $u_i$ can now be rewritten using this field
reparametrisation. For the particular choice of constants
\begin{eqnarray}
 {\bf c}_i &=& -4\m_i{\bf p}_i\\
 {\bf s}_i &=& \frac{1}{2}(\m_i+1){\bf p}_i\\
 {\bf q}_i &=& \m_i{\bf p}_i\; ,
\end{eqnarray}
\noindent where $\m_i\in\{ -1,+1\}$ are arbitrary signs, we find
that the Lagrangian simplifies to
\begin{equation}
 L = \sqrt{-g}\left( R+\frac{1}{2}\sum_A\pt_\m\f_A\pt^\m\f_A+8
     \sum_ie^{-{\bf P}_i\cdot\bph}\pt_\m z_i\pt^\m z_i\right) \label{L}
\end{equation}
where
\begin{equation}
 {\bf P}_i = -\m_i{\bf p}_i\; .
\end{equation}
The obvious advantage of this form is that the rescaled blow-up
moduli $z_i$ (unlike their counterparts $u_i$) only appear through
their derivatives. This implies that their equations of motion can
be immediately integrated once, leading to a set of first
integrals. We will use this property in the following section when
we discuss the cosmological evolution based on the
Lagrangian~\eqref{L}. Since we have neglected higher order terms
in $\e_{(\t ,n)}$ in its derivation, it is equivalent to the one
directly obtained from the K\"ahler potential~\eqref{K} only
provided fields are in the region of moduli space defined by
Eq.~\eqref{app2}. Also note that this equivalence holds for an
arbitrary choice of the signs $\m_i$.

\section{Cosmological evolution equations}

To study the cosmological evolution based on the Lagrangian~\eqref{L}
we start with a metric
\begin{equation}
 ds^2=-e^{2\n (\t )}d\t^2+e^{2\a (\t )}d{\bf x}^2
\end{equation}
of Friedmann-Robertson-Walker type, with the spatial sections taken
to be flat for simplicity, and time-dependent moduli fields
\begin{equation}
 \f_A=\f_A (\t )\; ,\qquad z_i=z_i(\t )\; .
\end{equation}
Here, $\a (\t )$ is the scale factor of the universe and $\n (\t
)$ is the lapse function, which we will determine later. From
Eq.~\eqref{L} the equations of motion for the blow-up moduli $z_i$
are given by
\begin{equation}
 \frac{d}{d\t}\left(e^{-{\bf P}_i\cdot\bph +3\a -\n}\dot{z}_i\right)=0\; ,
\end{equation}
where the dot denotes the derivative with respect to $\t$. They can be
easily integrated to
\begin{equation}
 \dot{z}_i = \z_i\, e^{{\bf P}_i\cdot\bph -3\a +\n}\; , \label{zsol}
\end{equation}
with arbitrary integration constants $\z_i$. Let us now choose the
particularly convenient gauge $\n = 3\a$. In this gauge, the sum
of the $(00)$ component and the spatial components of the Einstein
equations reads $\ddot{\a}=0$ and, we have
\begin{equation}
 \a = v\t +\a_0 \label{a0}
\end{equation}
where $v$ and $\a_0$ are arbitrary integration constants. Proper time $t$
is obtained by integrating
\begin{equation}
 dt = e^{3\a}d\t
\end{equation}
which leads to
\begin{equation}
 v\t +\a_0 = \frac{1}{3}\ln\left(\sqrt{3E}|t-t_0|\right)\; .
 \label{pt}
\end{equation}
Here $t_0$ is another integration constant and
\begin{equation}
 E =3v^2\; .
\end{equation}
The solution~\eqref{a0} for the scale factor, written in proper
time, then takes the form
\begin{equation}
 \a = \frac{1}{3}\ln\left(\sqrt{3E}|t-t_0|\right)
\end{equation}
and shows the expected power-law behaviour, with power $1/3$,
characteristic of evolution driven by kinetic energy. As
usual, we have two branches, the negative-time branch, $t-t_0<0$,
which ends in a future curvature singularity, and the
positive-time branch, $t-t_0>0$, which starts out in a past
curvature singularity.

\vspace{0.4cm}

We still have to consider the equations of motion for the bulk moduli $\f_A$.
Replacing the blow-up moduli with Eq.~\eqref{zsol} and working in the
gauge $\n =3\a$ they take the form
\begin{equation}
 \frac{1}{2}\frac{d^2\f_A}{d\t^2}+\frac{\pt V}{\pt\f_A} = 0\; . \label{feom}
\end{equation}
The ``effective" potential
\begin{equation}
 V = 4\sum_i\z_i^2\, e^{{\bf P}_i\cdot\bph} \label{V}
\end{equation}
in these equations of course originates from integrating out the blow-up
moduli. The above equations of motion can be obtained from the Lagrangian
\begin{equation}
 \cL = \frac{1}{4}\dot{\bph}\cdot\dot{\bph}-V\; . \label{Lf}
\end{equation}
They have to be supplemented by the Hamiltonian constraint
\begin{equation}
 \cH = \frac{1}{4}\dot{\bph}\cdot\dot{\bph}+V=E\; , \label{H}
\end{equation}
which is simply the Friedmann equation rewritten in our language.

\vspace{0.4cm}

For the subsequent discussion it is useful to express the quantities
measuring the geometry of the $G_2$ manifold in terms of the
redefined fields. The volume $\cV_0$ of the underlying orbifold, defined
in Eq.~\eqref{V0}, can then be written
as
\begin{equation}
 \cV_0 = \frac{1}{8}\exp\left(\frac{1}{3}\sum_A\f_A\right)\; ,
 \label{V1}
\end{equation}
where we have neglected higher order terms. Likewise, the fraction
$\e_{(\t ,n)}$ of the total volume taken up by the blow-up
$(\t ,n)$ now takes the form
\begin{equation}
 \e_{(\t ,n)} = \frac{8}{3}\sum_ae^{-{\bf P}_i\cdot\bph}\,
                z_{(\t ,n,a)}^2\; .
 \label{e1}
\end{equation}

\vspace{0.4cm}

Let us summarise what we have achieved so far. We have integrated
out the blow-up moduli and decoupled the scale factor by choosing
a particular gauge. This has reduced the problem of analysing the
evolution of $43$ freely rolling $G_2$ moduli to one that only
involves the seven bulk moduli subject to the potential~\eqref{V}.
The remaining problem is, therefore, to solve the seven equations
of motion~\eqref{feom} for the bulk moduli and the
constraint~\eqref{H}. Each solution for the bulk moduli can then
be inserted into Eq.~\eqref{zsol}, which determines the
corresponding evolution of the blow-up moduli.

There exists a well-developed solution theory~\cite{Kostant:qu},
\cite{Lu:1996hh}--\cite{Lukas:1996zq} for the
Lagrangian~\eqref{Lf} and the associated equations of
motion~\eqref{feom} for the bulk moduli if the system is of Toda type.
To discuss this more specifically we introduce the matrix
\begin{equation}
 A_{ij}={\bf P}_i\cdot{\bf P}_j\; , \label{A}
\end{equation}
which consists of one or more irreducible blocks along the diagonal.
The system is called Toda if each of these blocks is proportional
to the Cartan matrix of a simple Lie-group. One can now inspect
the vectors ${\bf P}_i=-\m_i{\bf p}_i$ from Table~1, where we recall
that the $\m_i$ are arbitrary signs. It is clear that the complete
nine-dimensional matrix obtained from~\eqref{A} is not proportional
to a Cartan matrix for any choices of the signs $\m_i$ and, hence,
the system is not of Toda type in this case. However, note that the
potential~\eqref{V} does not necessarily contain all possible terms.
In fact, each term is multiplied by an arbitrary integration constant
$\z_i$ that can be set to zero. Such a vanishing $\z_i$ implies,
from Eq.~\eqref{zsol}, that the corresponding blow-up modulus $z_i$
is constant. Hence, we learn that we can consistently freeze an
arbitrary subset of blow-up moduli $z_i$ and thereby select an
arbitrary subset of the vectors ${\bf P}_i$ that appear in the
potential~\eqref{V}. Then, for suitable subsets and choices of signs
$\m_i$, the associated matrix $A$ can well be proportional to
a Cartan matrix, as we will see for several explicit examples studied
further below. The system is then Toda and can be integrated
analytically. The resulting solutions are, of course,
special in that they rely on a number of blow-up moduli being
frozen. However, in the present context, this seems to be the best
one can do by analytic methods.

\vspace{0.4cm}

Before we go on to analyse explicit examples, it is worth making a
general observation about the structure of the solutions. It is
well-known~\cite{Lukas:1996ee,Lukas:1996iq}, for Lagrangians of
the type~\eqref{L}, that the fields $z_i$ approach constant values
asymptotically at early and late times (both in the negative and
the positive time branch). This behaviour will also be confirmed
in the explicit examples. Further note that the
potential~\eqref{V} consists of a sum of positive terms.
Therefore, one expects the modes ${\bf P}_i\cdot\bph$ to roll up
the exponential slope at early time, then turn around and roll
down at late time. In other words, the exponentials $\exp ({\bf
P}_i\cdot\bph )$ decrease both in the past and in the future, at
least if they explicitly appear in $V$, due to their associated
integration constants $\z_i$ being non-zero. Hence, for the moduli
that are not completely frozen, the associated relative volumes of
the blow-ups, $\e_{(\t ,n)}$, always increases asymptotically at
early and late times. We will confirm this general feature
explicitly in our examples.

%%%%%%%%%%%%%%%%%%%%%%%%%%%%%%%%%%%%%%%%%%%%%%%%%%%%%%%%%%%%%%%%%%%%%%%%%%%

\section{A universal solution}

As a warm-up, we would first like to consider a universal solution
where all bulk moduli and all blow-up moduli evolve in the same
way. It appears that a natural way to obtain such a solution is to
start with an Ansatz where the seven bulk moduli are equal, that
is where there is only a single breathing mode. However, it turns
out that such an Ansatz is incompatible with the equations of
motion~\eqref{feom}. The reason behind this is that not all bulk
moduli are on the same footing because of their differing couplings to
the blow-up moduli, as is evident from the coupling vectors in
Table~1. Instead, we start with the slightly more general Ansatz
\begin{equation}
 \f_A = c_A\f \label{A1}
\end{equation}
where $\f$ is the breathing mode and $c_A$ are constants to be determined.
For this Ansatz to be successful we need the seven equations of
motion~\eqref{feom} for $\f_A$ to reduce to a single equation for $\f$.
This leads to the conditions
\begin{eqnarray}
 {\bf p}_{(\t ,a)}\cdot{\bf c} &=& {\rm const} \label{c1} \\
 \sum_{\t ,a}\z_{(\t ,a)}\, p_{(\t ,a)A} &=& \frac{1}{2}c_A\z^2 \label{c2}
\end{eqnarray}
where we have defined
\begin{equation}
 \z_{(\t ,a)}^2 = \sum_n \z_{(\t ,n,a)}^2\; ,
\end{equation}
and $\z$ is a constant. These conditions lead to a unique solution
\begin{equation}
 {\bf c} = \frac{2}{7}(4,4,4,3,3,3,3)
\end{equation}
for the vector ${\bf c}$ in our Ansatz~\eqref{A1}. It can further be shown
that all conditions~\eqref{c1}, \eqref{c2} can be satisfied for
specific choices of coefficients $\z_{(\t ,a)}$ within a three-parameter
family of solutions. The details of this are inessential for our
subsequent discussion. For such a solution, the seven equations
of motion~\eqref{feom} reduce to the single equation
\begin{equation}
 \frac{1}{2}\ddot{\f}-2\m\z^2 e^{-2\m\f} = 0\; ,
\end{equation}
where $\m$ is an arbitrary sign. This equation can be easily integrated
once, and the integration constant can be fixed from the constraint~\eqref{H}.
This leads to the first integral
\begin{equation}
 E=|{\bf c}|^2\left(\frac{1}{4}\dot{\f}^2+\z^2e^{-2\m\f}\right)
\end{equation}
and the solution
\begin{equation}
 \f = \m\ln\cosh (y)
      +\frac{\m}{2}\ln\left(\frac{|{\bf c}|^2\z^2}{E}\right)
\end{equation}
for the breathing mode $\f$, where
\begin{equation}
 y=\frac{2\sqrt{E}}{|{\bf c}|^2}(\t -\t_1)
\end{equation}
is a rescaled time variable and $\t_1$ is a constant. The seven bulk moduli
are proportional to $\f$ and can be obtained by inserting this result
into Eq.~\eqref{A1}. From Eq.~\eqref{zsol} one can then, in turn,
obtain the solutions for the blow-up moduli. After another integration
one finds
\begin{equation}
 z_i = \frac{d_i}{2}\left(\tanh (y)+1\right)+z_{0i}\;
\end{equation}
where
\begin{equation}
 d_i = \frac{\sqrt{E}\z_i}{\z^2}
\end{equation}
and $z_{0i}$ are independent integration constants. As advertised earlier,
the blow-up moduli $z_i$ indeed approach constants asymptotically.
More precisely, at early time, $y\rightarrow -\infty$, we have
$z_i\rightarrow z_{0i}$ and at late time, $y\rightarrow\infty$, we have
$z_i\rightarrow z_{1i}\equiv z_{0i}+d_i$, that is, $d_i$ is the distance
by which the blow-up modulus $z_i$ moves. We also note that the non-trivial
evolution of $z_i$ happens around the time $y\simeq 0$.

\vspace{0.4cm}

Let us now interpret this solution. From Eq.~\eqref{V1} the orbifold
volume is directly measured by the breathing mode $\f$ via
\begin{equation}
 \cV_0 = \frac{1}{8}\exp\left(\frac{16}{7}\f\right)\; .
\end{equation}
Further, from Eq.~\eqref{e1}, the relative volume of a blow-up
is given by
\begin{equation}
 \e_i = \frac{8}{3}e^{2\m\f}z_i^2=\frac{2}{3}|{\bf c}|^2\frac{\z_i^2}{\z^2}
        \left(\frac{z_{0i}}{d_i}e^{-y}+\frac{z_{1i}}{d_i}e^y\right)^2\; .
\end{equation}
We recall that our solution can only be trusted as long as
$\e_i\ll 1$. Since, from Eq.~\eqref{c2}, the pre-factor on the RHS
of the above equation is of order one (for at least one $i$) we
have to require the term in bracket be smaller than one. This
leads to two cases, namely
\begin{itemize}
 \item $\e_i\ll 1$ if $|z_{0i}/d_i|\ll 1$ and $\ln |z_{0i}/d_i|\ll y\ll -1$.
       In this time range the breathing mode evolves as
       $\f\simeq -\m y +{\rm const}$.
 \item $\e_i\ll 1$ if $|z_{1i}/d_i|\ll 1$ and $1\ll y\ll\ln |d_i/z_{1i}|$.
       In this time range the breathing mode evolves as
       $\f\simeq \m y +{\rm const}$.
\end{itemize}
In both cases the breathing mode and, hence, the orbifold volume
$\cV_0$, can increase or decrease depending on the choice of the
sign $\m$. However, independent of this choice, the relative
blow-up volume $\e_i$ will always leave the allowed range $\e_i\ll
1$ when one of the limits of the given time ranges is approached.
This means that after a finite proper time, both in the past and in
the future, at least one of the blow-ups will take up a
significant portion of the space and the approximation on which
our K\"ahler potential~\eqref{K} is based brakes down. In
particular this means an evolution towards a state with small
blow-ups at late time, $y\rightarrow\infty$ is not possible.

%%%%%%%%%%%%%%%%%%%%%%%%%%%%%%%%%%%%%%%%%%%%%%%%%%%%%%%%%%%%%%%%%%%%%%%%%%

\section{Potential with a single exponential}

Let us now analyse the solutions more systematically, starting from
simple patterns of evolution of the blow-up moduli and moving to more
complicated ones.

\vspace{0.4cm}

Certainly, the simplest possibility is to freeze all blow-up moduli
by setting all $\z_i=0$ in Eq.~\eqref{zsol}. In this case, the effective
potential~\eqref{V} for the bulk moduli vanishes identically and
the equations of motion~\eqref{feom} for $\f_A$ can be easily
integrated. In proper time $t$, related to $\t$ by Eq.~\eqref{pt},
one easily finds as the general solution in this case
\begin{equation}
 \f_A = q_A\ln\left(\frac{|t-t_0|}{T}\right)+k_A\label{frees}
\end{equation}
where $t_0$, $T$ and $k_A$ are constants. The expansion
powers $q_A$ satisfy
\begin{equation}
 |{\bf q}|^2=\frac{4}{3}\; ,
\end{equation}
which follows from the constraint~\eqref{H}, and are otherwise arbitrary.
These are simply solutions describing power-law evolution of the
bulk moduli, which, in fact, are identical to the ones that can
be obtained for M-theory on a seven-dimensional torus.

\vspace{0.4cm}

We now move to the next more complicated case where blow-up moduli
of only one particular type $(\t ,a)$ evolve non-trivially and
all the others have been set to constants. We write ${\bf P}={\bf
P}_{(\t ,a)}$ and $\z^2=4\sum_n \z_{(\t ,n,a)}$ for simplicity of
notation. The effective potential consists of only one term and
takes the form
\begin{equation}
 V=\z^2 e^{{\bf P}\cdot\bph}\; .
\end{equation}
This situation corresponds to an $SU(2)$ Toda model so the general
solution to Eq.~\eqref{feom} can be found. It is given by
\begin{equation}
 \bph = {\bf p}^{(i)}\ln (x) +({\bf p}^{(f)}-{\bf p}^{(i)})
        \ln (1+x^\d )^{1/\d } + {\bf k} \label{singles}
\end{equation}
subject to the constraints
\begin{eqnarray}
 |{\bf p}^{(i)}|^2 &=& \frac{4}{3} \label{C1}\\
 \d &=& {\bf P}\cdot{\bf p}^{(i)} \label{C2}\\
 {\bf p}^{(f)} &=& {\bf p}^{(i)}-\frac{2{\bf P}\cdot{\bf p}^{(i)}}
                   {|{\bf P}|^2}{\bf P} \label{C3}\\
 \exp ({\bf P}\cdot{\bf k}) &=& \frac{3E\d^2}{|{\bf P}|^2\z^2}\; .
 \label{C4}
\end{eqnarray}
Further,
\begin{equation}
 x=\frac{|t-t_0|}{T}
\end{equation}
is the rescaled proper time and $t_0$ and $T$ are constants. Note that
Eqs.~\eqref{C1} and \eqref{C3} imply that
\begin{equation}
 |{\bf p}^{(f)}|^2= \frac{4}{3}\; .
\end{equation}
This solution is invariant under the exchange of ${\bf p}^{(i)}$ and
${\bf p}^{(f)}$ and we remove this ambiguity by requiring $\d\geq 0$
in the positive-time branch and $\d\leq 0$ in the negative-time
branch. The interpretation of these solutions is
well-known~\cite{Lukas:1996ee,Lukas:1996iq}.
They interpolate between two, generally different, ``free"
solutions~\eqref{frees}, one with ${\bf q}={\bf p}^{(i)}$ at early time
and one with ${\bf q}={\bf p}^{(f)}$ at late time. In these asymptotic
regions the blow-up moduli $z_{(\t ,n,a)}$ are constant while they move
around the time $x\simeq 1$ to facilitate the transition between the
two free solutions.

\vspace{0.4cm}

One can now compute the relative volume $\e_{(\t ,a)}$
of the blow-ups by inserting the solution~\eqref{singles} into Eq.~\eqref{e1}.
As before, we find that the required condition $\e_{(\t ,a)}\ll 1$
is satisfied only in two finite time windows at $x\ll 1$ and $x\gg 1$.
At the endpoints of these windows the volume of the blow-ups
becomes sizeable and control over the approximation is lost.

%%%%%%%%%%%%%%%%%%%%%%%%%%%%%%%%%%%%%%%%%%%%%%%%%%%%%%%%%%%%%%%%%%%%%%%%%%%%%%%

\section{More complicated cases}

An obvious generalisation of the previous example is to consider a situation
that corresponds to an $SU(2)^n$ Toda model, for some
integer $n$. This amounts to having a subset of blow-up moduli
evolve non-trivially for which the associated characteristic
vectors ${\bf P}_i$ are orthogonal, that is,
\begin{equation}
 {\bf P}_i\cdot {\bf P}_j = 2\d_{ij}\; . \label{ortho}
\end{equation}
Inspection of Table~1 shows such cases can indeed be realised. For
example, allowing the moduli of a certain type $\t$ to evolve
non-trivially, while moduli of the other two types are being kept
constant by setting $\z_{\t ',n,a}=0$ in Eq.~\eqref{zsol} for $\t
'\neq\t$, leads to an $SU(2)^3$ Toda model with three exponentials
appearing in the potential~\eqref{V}. From Table~1 there are a
number of other options and in the following we will simply assume
the existence of $n$ vectors ${\bf P}_i$ satisfying~\eqref{ortho}
to cover all possibilities.

\vspace{0.4cm}

The standard procedure to deal with such a system is to introduce a
new constant basis ${\bf e}_A=({\bf e}_i,{\bf e}_a)$ in field space, where
$i =1,\cdots ,n$ and $a=n+1,\cdots ,7$ such that
\begin{equation}
 {\bf e}_i = \frac{1}{\sqrt{2}}{\bf P}_i
\end{equation}
and the remaining vectors ${\bf e}_a$ are chosen to complete the set
to an orthonormal basis satisfying
\begin{equation}
 {\bf e}_A\cdot{\bf e}_B = \d_{AB}\; .
\end{equation}
The bulk fields $\bph$ can then be expanded as
\begin{equation}
 \bph = \sum_A\r_A{\bf e}_A\; ,
\end{equation}
where $\r_A$ is a new set of fields. The Lagrangian~\eqref{Lf} and
the Hamiltonian~\eqref{H}, written in terms of these new fields,
take the form
\begin{equation}
 \cL = \frac{1}{4}\sum_A\dot{\r}_A^2-V\; ,\qquad
 \cH = \frac{1}{4}\sum_A\dot{\r}_A^2+V=E\; , \label{LH}
\end{equation}
with the potential
\begin{equation}
 V = 4\sum_i\z_i^2e^{2\r_i}\; .
\end{equation}
Note that the modes $\r_a$ have decoupled from this potential,
which is, of course, one of the motivations behind introducing
the basis ${\bf e}_A$.

\vspace{0.4cm}

The general solution to the Lagrangian in~\eqref{LH} can be easily obtained as
\begin{equation}
 \bph = \sum_A\r_A{\bf e}_A
\end{equation}
with
\begin{eqnarray}
 \r_a &=& k_a\t +\t_a \\
 \r_i &=& \sqrt{2}\ln\cosh (y_i)
          -\frac{1}{\sqrt{2}}\ln\left(\frac{16\z_i^2}{k_i^2}\right)\; ,
\end{eqnarray}
where $\t_A$ and $k_A$ are constants and
\begin{equation}
 y_i = \frac{k_i}{\sqrt{2}}(\t -\t_i)
\end{equation}
are rescaled time coordinates. The Hamiltonian constraint
in~\eqref{LH} amounts to the condition
\begin{equation}
 \sum_Ak_A^2 = 4E\; .
\end{equation}
As expected the modes $\r_a$ evolve freely. Each of the other
modes $\r_i$ evolves similarly to what has been found for the
single $SU(2)$ Toda model in the previous section. This can be seen
explicitly by converting the above solution to proper time
using~\eqref{pt} and comparing with Eq.~\eqref{singles}. This
means that each mode interpolates between two regions of simple
power-law evolution at early and late time. The blow-up moduli
$z_i$ are constant in these asymptotic regions and their evolution
at intermediate times facilitates the transition. This can be
 seen explicitly from their solution
\begin{equation}
 z_i=\frac{d_i}{2}\left(\tanh (y_i)+1\right)+z_{0i}\; ,
\end{equation}
with $z_{0i}$ and $d_i$ constants representing the value of the
modulus at early time and its total change respectively.
This solution is obtained by inserting the above solution for
$\bph$ into Eq.~\eqref{zsol}.

From Eq.~\eqref{e1} one finds the relative volume
\begin{equation}
 \e_i = \frac{\sqrt{2}}{3}\left(\frac{z_{1i}}{d_i}e^{y_i}+\frac{z_{0i}}{d_i}
        e^{-y_i}\right)^2
\end{equation}
of the blow-ups. As in all previous cases, $\e_i$ can only be kept
small, $\e_i\ll 1$, for a finite proper time at either
$y_i\ll -1$ or $y_i\gg 1$. For a valid solution, all $\e_i$ need to
be small, which amounts to arranging an overlap between those regions
by choosing the time shifts $\t_i$ appropriately. Outside this overlap
region control of the approximation is lost as one or more blow-ups
become large and take up a significant part of the internal space.

\vspace{0.4cm}

There are also cases leading to a Toda model associated with a
higher-rank simple group. Consider, for example, the three
vectors
\begin{eqnarray}
 {\bf P}_{(\a ,1)} &=& (-1,0,0,0,0,-1,0) \label{P1}\\
 {\bf P}_{(\b ,1)} &=& (1,0,0,0,0,0,1) \label{P2}\\
 {\bf P}_{(\g ,3)} &=& (0,-1,0,0,0,0,-1)\; ,\label{P3}
\end{eqnarray}
obtained from Table~1 with the sign choice $\m_{(\a ,1)}=-1$,
$\m_{(\b ,2)}=1$ and $\m_{(\g ,3)}=-1$ and set all blow-up
moduli with $(\t ,a)$ different from the above to constants.
The matrix $A$ in Eq.~\eqref{A} is then given by
\begin{equation}
 (A_{\t\s}) = \left(\begin{array}{rrr}2&-1&0\\-1&2&-1\\0&-1&2\end{array}\right)\; ,
\end{equation}
which is precisely the Cartan matrix of $SU(4)$. Hence, we are dealing
with an $SU(4)$ Toda model.

As before, the first step in solving this model is to introduce a new
basis $({\bf e}_A)=({\bf e}_\t ,{\bf e}_a)$, where $\t = 1,2,3$ and
$a=4,\dots ,7$, in field space. Here the three vectors ${\bf e}_\t$
are identified with~\eqref{P1}--\eqref{P3} and the remaining
four vectors ${\bf e}_a$ are chosen to be orthonormal among themselves
and orthogonal to all ${\bf e}_\t$. Consequently, we have a basis
satisfying
\begin{equation}
 {\bf e}_\t\cdot{\bf e}_{\s} = A_{\t\s}\; ,\qquad
 {\bf e}_\t\cdot{\bf e}_a = 0\; ,\qquad
 {\bf e}_a\cdot{\bf e}_b = \d_{ab}\; .
\end{equation}
Expanding the fields $\bph$
\begin{equation}
 \bph = \sum_A\r_A{\bf e}_A
\end{equation}
as before one finds for the Lagrangian~\eqref{Lf}
\begin{equation}
 \cL = \frac{1}{4}\sum_{\t ,\s}A_{\t ,\s}\dot{\r}_\t\dot{\r}_\s
       +\frac{1}{4}\sum_a\dot{\r}_a^2-V
\end{equation}
with potential
\begin{equation}
 V = 4\sum_\t\z_\t^2\,\exp\left(\sum_\s A_{\t\s}\r_\s\right)\; .
\end{equation}
The modes $\r_a$ are decoupled from this potential and their
equations of motion immediately lead to the general solution
\begin{equation}
 \r_a = k_a(\t -\t_a )\; ,
\end{equation}
where $k_a$ and $\t_a$ are constants. Following the methods of
Ref.~\cite{Kostant:qu}, the solutions for the other modes are obtained
as
\begin{equation}
 e^{-\r_\t} = \sum_{\bla\in\L_\t}b_\t (\bla )\exp\left(\bla\cdot ({\bf k}
              \t - \btau )\right)\; , \label{sol4}
\end{equation}
where ${\bf k}$ and $\btau$ are constant vectors. The three sets $\L_\t$
contain the weights of the fundamental representation ${\bf 4}$, the vector
representation ${\bf 6}$ and the anti-fundamental representation
$\bar{\bf 4}$ of $SU(4)$, respectively, and are explicitly given by
\begin{eqnarray}
 \L_1 &=& \{ (100),(-110),(0-11),(00-1)\} \\
 \L_2 &=& \{ (010),(1-11),(-101),(10-1),(-11-1),(0-10)\} \\
 \L_3 &=& \{ (001),(01-1),(1-10),(-100)\} \; .
\end{eqnarray}
Finally, $b_\t (\bla )$ represents a set of constants which depends
on $\z_\t$ as well as on ${\bf k}$. They can be calculated
by inserting the solutions~\eqref{sol4} into the equations of motion
but their explicit form will not be of any relevance here.

The important feature of~\eqref{sol4} is that asymptotically one
of the exponentials in the sum will dominate leading to a power-law
evolution. In these regions, the blow-up moduli are approximately
constant and the relative blow-up volumes increase until they
approach values of order one where the approximation underlying
our analysis breaks down. Hence, this more complicated Toda
solution also conforms with our general expectation of asymptotically
increasing blow-up volumes.

%%%%%%%%%%%%%%%%%%%%%%%%%%%%%%%%%%%%%%%%%%%%%%%%%%%%%%%%%%%%%%%%%%%%%%

\vspace{1cm}

\noindent
{\Large\bf Acknowledgments}\\
A.~L.~is supported by a PPARC Advanced Fellowship, and S.~M.~by a PPARC
Postgraduate Studentship.\\

%%%%%%%%%%%%%%%%%%%%%%%%%%%%%%%%%%%%%%%%%%%%%%%%%%%%%%%%%%%%%%%%%%%%%%%%%%%%


\begin{thebibliography}{01}

\bibitem{Papadopoulos:1995da}
G.~Papadopoulos and P.~K.~Townsend,
``Compactification of D = 11 supergravity on spaces of exceptional holonomy,''
Phys.\ Lett.\ B {\bf 357} (1995) 300
[arXiv:hep-th/9506150].
%%CITATION = HEP-TH 9506150;%%
\bibitem{Acharya:1998pm}
B.~S.~Acharya, ``M theory, Joyce orbifolds and super Yang-Mills,''
Adv.\ Theor.\ Math.\ Phys.\  {\bf 3} (1999) 227
[arXiv:hep-th/9812205].
%%CITATION = HEP-TH 9812205;%%
\bibitem{Gukov:1999gr}
S.~Gukov, ``Solitons, superpotentials and calibrations,'' Nucl.\
Phys.\ B {\bf 574} (2000) 169 [arXiv:hep-th/9911011].
%%CITATION = HEP-TH 9911011;%%
\bibitem{Atiyah:2001qf}
M.~Atiyah and E.~Witten,
``M-theory dynamics on a manifold of G(2) holonomy,''
Adv.\ Theor.\ Math.\ Phys.\  {\bf 6} (2003) 1
[arXiv:hep-th/0107177].
%%CITATION = HEP-TH 0107177;%%

\bibitem{Acharya:2000ps}
B.~S.~Acharya and B.~Spence,
``Flux, supersymmetry and M theory on 7-manifolds,''
arXiv:hep-th/0007213.
%%CITATION = HEP-TH 0007213;%%
%\cite{Cvetic:2001nr}
\bibitem{Cvetic:2001nr}
M.~Cvetic, G.~Shiu and A.~M.~Uranga, ``Chiral four-dimensional N =
1 supersymmetric type IIA orientifolds from  intersecting
D6-branes,'' Nucl.\ Phys.\ B {\bf 615} (2001) 3
[arXiv:hep-th/0107166].
%%CITATION = HEP-TH 0107166;%%
\bibitem{Witten:2001uq}
E.~Witten,
``Anomaly cancellation on G(2) manifolds,''
arXiv:hep-th/0108165.
%%CITATION = HEP-TH 0108165;%%

\bibitem{Acharya:2001gy}
B.~Acharya and E.~Witten,
``Chiral fermions from manifolds of G(2) holonomy,''
arXiv:hep-th/0109152.
%%CITATION = HEP-TH 0109152;%%

\bibitem{Witten:2001bf}
E.~Witten,
``Deconstruction, G(2) holonomy, and doublet-triplet splitting,''
arXiv:hep-ph/0201018.
%%CITATION = HEP-PH 0201018;%%

\bibitem{Beasley:2002db}
C.~Beasley and E.~Witten,
``A note on fluxes and superpotentials in M-theory compactifications on  manifolds of G(2) holonomy,''
JHEP {\bf 0207} (2002) 046
[arXiv:hep-th/0203061].
%%CITATION = HEP-TH 0203061;%%

\bibitem{Lazaroiu:2002jv}
C.~I.~Lazaroiu and L.~Anguelova, ``M-theory compactifications on
certain 'toric' cones of G(2) holonomy,'' JHEP {\bf 0301} (2003)
066 [arXiv:hep-th/0204249].

\bibitem{Anguelova:2002dd}
L.~Anguelova and C.~I.~Lazaroiu, ``M-theory on 'toric' G(2) cones
and its type II reduction,'' JHEP {\bf 0210} (2002) 038
[arXiv:hep-th/0205070].
%%CITATION = HEP-TH 0205070;%%



%%CITATION = HEP-TH 0204249;%%
\bibitem{Berglund:2002hw}
P.~Berglund and A.~Brandhuber,
``Matter from G(2) manifolds,''
Nucl.\ Phys.\ B {\bf 641} (2002) 351
[arXiv:hep-th/0205184].
%%CITATION = HEP-TH 0205184;%%
%\cite{Behrndt:2002xm}
\bibitem{Behrndt:2002xm}
K.~Behrndt, G.~Dall'Agata, D.~Lust and S.~Mahapatra,
``Intersecting 6-branes from new 7-manifolds with G(2) holonomy,''
JHEP {\bf 0208} (2002) 027 [arXiv:hep-th/0207117].
%%CITATION = HEP-TH 0207117;%%

\bibitem{Friedmann:2002ty}
T.~Friedmann and E.~Witten,
``Unification scale, proton decay, and manifolds of G(2) holonomy,''
arXiv:hep-th/0211269.
%%CITATION = HEP-TH 0211269;%%

\bibitem{Acharya:2002kv}
B.~S.~Acharya,
``A moduli fixing mechanism in M theory,''
arXiv:hep-th/0212294.
%%CITATION = HEP-TH 0212294;%%

%\cite{Lukas:2003dn}
\bibitem{Lukas:2003dn}
A.~Lukas and S.~Morris, ``Moduli K\"ahler potential for M-theory
on a G(2) manifold,'' arXiv:hep-th/0305078.
%%CITATION = HEP-TH 0305078;%%

\bibitem{joyce1}
D.~Joyce,
``Compact Riemannian 7-Manifolds with Holonomy $G_2$. I,''
J.\ Diff.\ Geom.\ {\bf 43} (1996) 291.

\bibitem{joyce2}
D.~Joyce,
``Compact Riemannian 7-Manifolds with Holonomy $G_2$. II,''
J.\ Diff.\ Geom.\ {\bf 43} (1996) 329.

\bibitem{joyceb}
D.~Joyce,
``Compact Manifolds with Special Holonomy'', Oxford Mathematical
Monographs, Oxford University Press, Oxford 2000.

%\cite{Kostant:qu}
\bibitem{Kostant:qu}
B.~Kostant,
``The Solution To A Generalized Toda Lattice And Representation Theory,''
Adv.\ Math.\  {\bf 34} (1979) 195.
%%CITATION = ADMTA,34,195;%%

%\cite{Lu:1996hh}
\bibitem{Lu:1996hh}
H.~Lu, C.~N.~Pope and K.~W.~Xu,
``Liouville and Toda Solutions of M-theory,''
Mod.\ Phys.\ Lett.\ A {\bf 11} (1996) 1785
[arXiv:hep-th/9604058].
%%CITATION = HEP-TH 9604058;%%

%\cite{Lu:1996jr}
\bibitem{Lu:1996jr}
H.~Lu and C.~N.~Pope,
``SL(N+1,R) Toda solitons in supergravities,''
Int.\ J.\ Mod.\ Phys.\ A {\bf 12} (1997) 2061
[arXiv:hep-th/9607027].
%%CITATION = HEP-TH 9607027;%%

%\cite{Lukas:1996ee}
\bibitem{Lukas:1996ee}
A.~Lukas, B.~A.~Ovrut and D.~Waldram,
``Cosmological solutions of type II string theory,''
Phys.\ Lett.\ B {\bf 393} (1997) 65
[arXiv:hep-th/9608195].
%%CITATION = HEP-TH 9608195;%%

%\cite{Lukas:1996iq}
\bibitem{Lukas:1996iq}
A.~Lukas, B.~A.~Ovrut and D.~Waldram,
``String and M-theory cosmological solutions with Ramond forms,''
Nucl.\ Phys.\ B {\bf 495} (1997) 365
[arXiv:hep-th/9610238].
%%CITATION = HEP-TH 9610238;%%

%\cite{Lukas:1996zq}
\bibitem{Lukas:1996zq}
A.~Lukas, B.~A.~Ovrut and D.~Waldram,
``Stabilizing dilaton and moduli vacua in string and M-theory cosmology,''
Nucl.\ Phys.\ B {\bf 509} (1998) 169
[arXiv:hep-th/9611204].
%%CITATION = HEP-TH 9611204;%%

\end{thebibliography}
\end{document}